\documentclass[a4paper, 10pt, conference]{ieeeconf}

\IEEEoverridecommandlockouts

\usepackage{dsfont}

\usepackage{graphics} % for pdf, bitmapped graphics files
\usepackage{epsfig} % for postscript graphics files
\usepackage{times} % assumes new font selection scheme installed
\usepackage{amsmath} % assumes amsmath package installed
\usepackage{amssymb}  % assumes amsmath package installed

\def\beq{\begin{equation}}
\def\eeq{\end{equation}}

\def\1{\mathbf{1}}

\def\Z{\mathbb{Z}}

\def\R{\mathbb{R}}

\def\Tr{{\rm Tr}\ }

\newtheorem{lemma}{Lemma}[section]
\newtheorem{corollary}{Corollary}[section]
\newtheorem{proposition}{Proposition}[section]
\newtheorem{example}{Example}

\title{\LARGE \bf
Optimal strategies in the average consensus problem}

\author{Jean-Charles Delvenne \and Ruggero Carli \and Sandro Zampieri\thanks{J.-C.
Delvenne is with the Institute for Mathematical Sciences, Imperial
College, 53 Prince's Gate, London SW7 2PG, United Kingdom,
{\tt\small jc.delvenne@imperial.ac.uk}. R. Carli and S. Zampieri are
with the Department of Information Engineering, Universit\`a di
Padova, Via Gradenigo 6/a, 35131 Padova, Italy {\tt\small
\{carlirug|zampi\}@dei.unipd.it}.}}

\begin{document}

 \maketitle \thispagestyle{empty}
\pagestyle{empty}
\begin{abstract}
We prove that for a set of communicating agents to compute the
average of their initial positions (average consensus problem), the
optimal topology of communication is given by a de Bruijn's graph.
Consensus is then reached in a finitely many steps. A more general
family of strategies, constructed by block Kronecker products, is
investigated and compared to Cayley strategies.
\end{abstract}

\section{Introduction}

Coordination algorithms for multiple autonomous vehicles and
decentralized estimation techniques for handling data coming from
distributed sensor networks have attracted large attention in recent
years. This is mainly motivated by that both coordinated control and
distributed estimation have applications in many areas, such as
coordinated flocking of mobile vehicles
\cite{Tanner+03a,Tanner+03b}, cooperative control of unmanned air
and underwater vehicles ~\cite{Leonard02, Beard01}, multi-vehicle
tracking with limited sensor information \cite{Speranzon},
monitoring very large scale areas with fine resolution
and collaborative estimation over wireless sensor networks \cite{Speranzon:06}.\\
Typically, both in coordinated control and in distributed estimation
the agents need to communicate data in order to execute a task. In
particular they may need to agree on the value of certain
coordination state variables. One expects that, in order to achieve
coordination, the variables shared by the agents, converge to a
common value, asymptotically. The problem of designing controllers
that lead to such asymptotic coordination is called
\emph{coordinated consensus}, see for
example~\cite{Fax04-1,Olfati04-1,Morse03-1} and references therein.
Generalisation  to high order consensus~\cite{RenMooreChen06} and nonholonomic
agents~\cite{LinFrancisMaggiore04, DimarogonasKyriakopoulos07, Tanner03}
have also been explored. %JCD: NEW SENTENCE ABOVE
%The interest in this type of problems is not limited to the field of
%mobile vehicles coordination but also in the field of
%synchronization theory~\cite{Strogatz,Marodi}.
One of the simplest consensus problems that has been mostly studied
consists in starting from systems described by an integrator and in
finding a feedback control yielding consensus, namely driving all
the states to the same value \cite{Olfati04-1}. The information
exchange is modeled by a directed graph describing in which pair of
agents the data transmission is allowed. The situation mostly
treated in the literature is when each agent has the possibility of
communicate its state to the other agents that are positioned inside
a neighborhood \cite{Tanner+03a,Morse03-1} and the communication
network is time-varying \cite{Tanner+03b,Morse03-1}. Robustness to
communication link failure \cite{Bullo} and the effects of time
delays \cite{Olfati04-1} has been considered recently. Randomly
time-varying networks have also been analyzed in \cite{Hatano}.
Moreover, a first analysis involving quantized data transmission has
been proposed in \cite{Zampieri06-1,Kashyap06-1}. In this paper we
consider the consensus problem from a different perspective. We are
interested to characterize the relationship between the amount of
information exchanged by the agents and the achievable control
performance. More precisely we assume that $N$ agents are given
initial positions in the euclidean space, and move in discrete-time
in order to reach the average of their initial positions. This
problem is also called \emph{average coordinated consensus}. Every
agent asks several agents their position  before taking a decision
to modify its own position. We impose that, in order to limit costs
of communication, every agent communicates with only $\nu$ agents
(including itself), where $\nu < N$. This means that in the graph
describing the communications between agents, the max in-degree is
at most $\nu$.
%This problem can be interpreted as vehicles achieving rendez-vous,
%or sensors measuring a same quantity and computing the average of
%their measurements.
%We will be particularly interested in linear time-invariant
%strategies to solve this problem, i.e., the vector of all positions
%at time $t+1$ is a linear function of the vector positions at time
%$t$.
%In [CarliZampiFagnaSpera], strategies based on Cayley communications
%graphs have been investigated.
In this paper, we exhibit a family of strategies for solving this
problem based on de Bruijn's graphs and we prove that according to a
suitable criteria this is the best that one can do. Precisely we
compute its performances according two criteria: rate of convergence
to the average of their initial positions and an LQR criterion. We
find that a deadbeat strategy is optimal according to the rate of
convergence, and nearly optimal according to the LQR criterion.
Finally, we compare it with an another strategy having limited
communication and exhibiting symmetries: the Cayley strategies
\cite{Zampieri05,Carli}. It should be noted however that our strategy is
limited to the case where the number of agents is an exact power of $\nu$.
Whether it is possible to build a linear time-invariant deadbeat strategy
for any number of agents (for a given $\nu$) remains an open problem. 

%JCD:THE TWO SENTENCES ABOVE ARE NEW.

The paper is organized as follows. In
Section II we provide some basic notions of graph theory and some
notational conventions. In Section III we formally define the
average consensus problem. In Section IV we introduce the block
Kronecker strategy. In Section V we show that the block Kronecker
strategy is the quickest possible strategy and we compare it with
the Cayley strategy. In section VI we evaluate the performance of
the block Kronecker strategy according to suitable quadratic
criteria.
 Finally we gather our conclusions in Section VII.
%%%
%%%Agents $1,\ldots, N$ are connected with respect to a graph. The
%%%average degree is fixed and presumably low. A strategy is given by a
%%%stochastic matrix. If we are to reach the barycentre, then the
%%%matrix has to be doubly stochastic.
%%%
%%%We compare three strategies. The first is a Cayley strategy, which
%%%is based on a Cayley graph of an abelian group; it is studied in
%%%[CarliZampiFagnaSpera]. The second one is a strategy the we
%%%introduce in this paper. The third one is deadbeat???

\section{Preliminaries on graph theory}

Before defining the problem we want to solve, we summarize some
notions on graph theory that will be useful throughout the rest of
the paper.

Let $\mathcal{G}=(V,\mathcal{E})$ be a directed \emph{graph} where
$V=(1,\ldots,N)$ is the set of \emph{vertices} and $\mathcal{E}
\subset V\times V$ is the set of \emph{arcs} or \emph{edges}. If
$(i,j) \in \mathcal{E}$ we say that the arc $(i,j)$ is outgoing from
$i$ and incoming in $j$. The \emph{adjacency matrix} $A$ is a
$\left\{0,1\right\}$-valued square matrix indexed by the elements in
$V$ defined by letting $A_{ij}=1$ if and only if
$(i,j)\,\in\,\mathcal{E}$. Define the \emph{in-degree} of a vertex
$j$ as $\sum_iA_{ij}$ and the \emph{out-degree} of a vertex $i$ as
$\sum_jA_{ij}.$ In our setup we admit the presence of self-loops. A
graph is called \emph{in-regular} (\emph{out-regular}) of degree $k$
if each vertex has in-degree (out-degree) equal to $k$. A
\emph{path} in ${\cal G}$ consists of a sequence of vertices
$i_1i_2\ldots\ldots i_r$ such that $(i_\ell,i_{\ell+1})\in
\mathcal{E}$ for every $\ell=1,\dots ,r-1$; $i_1$ (resp. $i_r$) is
said to be the \emph{initial} (resp. \emph{terminal}) vertex of the
path. A \emph{cycle} is a path in which the initial and the terminal
vertices coincide. A vertex $i$ is said to be \emph{connected} to a
vertex $j$ if there exists a path with initial vertex $i$ and
terminal vertex $j$. A directed graph is said to be \emph{connected}
if, given any pair of vertices $i$ and $j$, either $i$ is connected
to $j$ or $j$ is connected to $i$. A directed graph is said to be
\emph{strongly connected} if, given any pair
of vertices $i$ and $j$, $i$ is connected to $j$.\\
Finally some notational conventions. Let $A$ any matrix belonging to
$\R^{N\times N}$. With $\Tr A$ we denote the trace of $A$, i.e. the
sum of the diagonal entries. We say that $A$ is nonnegative, denoted $A\geq0$,
or positive, denoted $A>0$, if the entries of $A$ are respectively nonnegative
or positive.

\section{Problem formulation} \label{section3}

We suppose that the positions of all $N$ agents are listed into one
vector of dimension $N$. If the agents move, say, in $\R^3$, it
seems that we would need a $3N$-dimensional vector. However we will
suppose that the positions are scalar, as every linear strategy on
scalar positions, if applied separately on every component of the
position, trivially extends to strategies for higher dimensions.

More precisely the problem of our interest can be formalized in the
following way. Consider $N>1$ identical systems whose dynamics are
described by the following discrete time state equations
$$
    x_i^+=x_i+u_i\qquad i=1,\ldots,N
$$
where $x_i\in\R$ is the state of the $i$-th system, $x_i^+$
represents the updated state and $u_i\in\R$ is the control input.
More compactly we can write
\begin{equation}\label{stateeq}
    x^+=x+u
\end{equation}
where $x,u\in\R^N$. The goal is to design a feedback control law
$$
u=Kx,\qquad K\,\in\,\R^{N \times N}
$$
yielding the average consensus, namely a control such that all the
$x_i$'s become asymptotically equal to the average of the initial
condition. More precisely, our objective is to obtain $K$ such that,
for any initial condition $x(0)\,\in\,\R^N$, the closed loop system
$$
x^{+}=(I+K)x,
$$
yields
\begin{equation}\label{agree}
\lim_{t\rightarrow\infty}x(t)=\alpha \1
\end{equation}
where $\1:=[1,\ldots,1]^T$ and
\begin{equation}\label{alpha}
\alpha=\frac{1}{N}\1^Tx(0).
\end{equation}

%If we restrict to control laws $K$ making $I+K$ nonnegative, namely a matrix %with all the components nonnegative, it is well known in the literature %that condition (\ref{agree}) and (\ref{alpha}) impose that $I+K$ is a doubly %stochastic matrix.

Writing $x(t)$ as a linear combination of the eigenvectors of $I+K$, 
%JCD: "Writing..." IS NEW
it is almost immediate to see that the average consensus problem is solved
if and only if the following three conditions hold:
\begin{itemize}
\item[(A)] Every row and every column of $I+K$ sums to one. Hence it has
eigenvalue $1$ with $\1$ as left and right eigenvector.
\item[(B)] The eigenvalue $1$ of $I+K$ has algebraic multiplicity one
(namely it is a simple root of the characteristic polynomial of
$I+K$).
\item[(C)] All the other
eigenvalues are strictly inside the unit circle.
%\item[(D)] Every row of $I+K$ contains at most  $\nu$ non-zero elements.
\end{itemize}

For nonnegative matrices, namely for matrices having all the
components nonnegative, condition (A) is called double
stochasticity, condition (B) is ergodicity and condition (C) is a
consequence of double stochasticity. We do not require our matrices
to be nonnegative, even though it will appear that the optimal
matrices are.

Observe now that the fact that the element in position $i,j$ of the
matrix $I+K$ is different from zero, means that the system $i$ needs
to know exactly the state of the system $j$ in order to compute its
feedback action. This implies that the $j$-th agent  must
communicate his state $x_j$ to $i$-th agent. In this context a good
description of the communication effort required by a specific
feedback $K$ is given by the directed graph ${\cal G}_{I+K}$ with
set of vertices $\{1,\dots ,N\}$ in which there is an arc from $j$
to $i$ whenever in the feedback matrix $K$ the element
$(I+K)_{ij}\not=0$. The graph ${\cal G}_K$ is said to be the
\emph{communication graph} associated with $K$. Conversely, given
any directed graph ${\cal G}$ with set of vertices $\{1,\dots ,N\}$,
a feedback $K$ is said to be {\it compatible} with ${\cal G}$ if
${\cal G}_{I+K}$ is a subgraph of ${\cal G}$ (we will use the
notation ${\cal G}_{I+K}\subseteq {\cal G}$).

In the sequel, we will impose the following constraint on the
communication graph: the max in-degree of the nodes is $\nu$. This
models the fact that communication lines are costly to establish or
operate, and every agent has the right to talk to a limited number
of other agents. Note that for compatibility with usual conventions
we consider that  $\nu$ counts all arcs entering a node, including
self-loops (which could be considered as `free communication' in
most technological situations).

Without this constraint, the problem becomes trivial: choose the
complete graph, and the consensus is reached in one step. We
therefore add the following constraint on $I+K$:

\begin{itemize}
\item[(D)] Every row of $I+K$ contains at most  $\nu$ non-zero elements.
\end{itemize}

From this point of view we would like to obtain a matrix $I+K$
satisfying (A),(B),(C),(D) and minimizing a suitable performance
index. The simplest control performance index is the exponential
rate of convergence to the average consensus. When we are dealing
with average consensus controllers it is meaningful to consider the
displacement from the average of the initial condition
$$
    \Delta(t):=x(t)-\left(\frac{1}{N}\1^Tx(0)\right) \1
\,.
$$
It is immediate to check that,
$\Delta(t)=x(t)-\left(\frac{1}{N}\1^Tx(t)\right) \1$ (since the average position
$\frac{1}{N}\1^Tx(t)$ is the same at all times $t$) %JCD: MODIFIED SENTENCE
and that it
satisfies the closed loop equation
\begin{equation}
    \label{eq:sameloop}
    \Delta^+=(I+K)\Delta\,.
\end{equation}
Notice moreover that the initial conditions $\Delta(0)$ are such
that
\begin{equation}
    \label{eq:orthog}
    \1^T\Delta(0)=0\,.
\end{equation}
Hence the asymptotic behavior of our consensus problem can
equivalently be studied by looking at the
evolution~\eqref{eq:sameloop} on the hyperplane characterized by the
condition~\eqref{eq:orthog}. The speed of convergence toward the
average of the initial condition can be defined as follows. Let $P$
any matrix satisfying conditions (A),(B),(C). Define
\begin{align*}
    \label{eq:spectral_rad}
    \rho(P)=
    \left\{\begin{array}{ll}1 &{\rm if}\;\; \dim\ker(P-I)>1\\
    \max_{\lambda \in\sigma(P)\setminus\{1\}}|\lambda|&{\rm
    if}\;\;
    \dim \ker(P-I)=1\,,\end{array}\right.
\end{align*}
which is called the \emph{essential spectral radius} of $P$. As the dominant
eigenvalues of $P^t$ is one and the others are smaller in magnitude than
$\rho(P)^t$, the essential spectral radius says how quickly $P^t$ converges
to the rank-one matrix $1/N \1 \1^T$, where $N$ is the dimension of $P$.
%JCD:NEW SENTENCE ABOVE
In this context the index $\rho(I+K)$ seems quite appropriate for analyzing
how performance is related to the communication effort associated
with a graph. The smaller the essential spectral radius, the quicker
the system will converge to the average of the initial condition.

However in control theory, strategies that converge in finite time
or  very quickly are sometimes dismissed on the ground that they
lead to large values of update values $u(t)=x(t+1)-x(t)$, that can
be physically impossible or very costly to implement. Hence a
strategy is often required to optimize an \emph{LQR cost}, taking
into account both the quickness convergence and the norm of updates
values. Therefore another suitable measure of performance could be
the following quantity:

\begin{equation}\label{cost}
J=\mathbb{E}(\sum_{t \geq 0} ||x(t)-x(\infty)||^2 + \gamma
||u(t)||^2),
\end{equation}
where $x(t)$ is the vector of positions at time $t$, $x(\infty)=\lim_{\infty}
x(t)$ %JCD: \lim_\infty x(t)  is new
is the vector whose every entry is the average of initial positions,
$u(t)=x(t+1)-x(t)$ is the update vector at time $t$, the initial
positions are supposed to be uncorrelated random variables with unit
variance, $\mathbb{E}$ denotes the expectation, $||x||^2=x^T x$ is
the euclidean norm and $\gamma$ is a nonnegative real.

We will prove that the optimal topology of communication (in the
meaning of speed of convergence) is given by a de Brujin's graph. We
will call the control strategies based on such graph block Kronecker
strategies, as explained in the next section. For these strategies
we will evaluated (\ref{cost}) and we will compare them to another
family of strategies based on a regular communication graph having
the same degree $\nu$: the Cayley
strategies~\cite{Zampieri05,Carli}.

\section{Block Kronecker strategies}

In this section, we define block Kronecker strategies.
Let $A$ be a $n \times n$  matrix satisfying (A),(B),(C),(D) and $k$
be a nonnegative integer. Note that if $A$ is full then $n \leq
\nu$ (since the number of non-zero elements cannot exceed $\nu$). Then we
build an $n^k \times n^k$ matrix $M$ in the following way. Let
$$
A=\left[
\begin{array}{c}
a_0\\
a_1\\
\vdots\\
a_{n-1}
\end{array}
\right]
$$
be a row-partition of the matrix $A$, where $a_i\,\in\,\R^{1\times
n}$. Then $M$ is the matrix
\begin{equation}\label{eq:M}
M= \left[
\begin{array}{c}
I_{n^{k-1}} \otimes a_0\\
I_{n^{k-1}} \otimes a_1\\
\vdots\\
I_{n^{k-1}} \otimes a_{n-1}
\end{array}
\right].
\end{equation}
For example, if $$A=\begin{pmatrix} \alpha & \beta \\ \beta & \alpha
\end{pmatrix}$$ (with $\alpha+\beta=1$) and $k=3$, then
$$M= \begin{pmatrix} \alpha & \beta &  &  & & & & \\
 & & \alpha & \beta &  &  & &   \\
 & & & & \alpha & \beta &  &  \\
 & & & & & & \alpha & \beta \\
 \beta & \alpha  &  &  & & & & \\
 & & \beta & \alpha  &  &  & &   \\
 & & & & \beta & \alpha  &  &  \\
 & & & & & &\beta & \alpha
  \end{pmatrix}$$

This is a kind of block Kronecker product. A general theory of block
Kronecker product is built in \cite{Koning91}. We only need a more
restricted definition, detailed below. The new matrix $M$ is a
matrix of larger dimension than $A$ and satisfying conditions
(A),(B),(C),(D): (A) and (D) follow from the definition, while (B)
and (C) are proved below. Hence it can play the role of the matrix $I+K$
in Section \ref{section3}.%JCD:NEW SENTENCE
We start by some reminders on Kronecker
product, define the block Kronecker product and explore the
properties of the latter.

\subsection{Kronecker product}
We recall that the \emph{Kronecker product} $A \otimes B$ of the
matrices $A$ and $B$ is the matrix $[a_{ij}B]_{i,j}$, whose
dimensions are the product of dimensions of $A$ and $B$. Some useful
properties of the Kronecker product are the following:
\begin{itemize}
\item $AB \otimes CD= (A \otimes C)(B \otimes D)$;

\item $\Tr A\otimes B=\Tr A \Tr B$;

\item the eigenvalues of $A\otimes B$ are all possible products of
an eigenvalue of $A$ with an eigenvalue of $B$;

\item  the eigenvectors of $A\otimes B$ are all possible Kronecker products
of an eigenvector of $A$ with an eigenvector of $B$.

\end{itemize}

The Kronecker product is sometimes called tensor product. Let us see
why. For instance consider the  matrices $B,C,D$ of sizes $m_B
\times n_B$, $m_C \times n_C$, $m_D \times n_D$. The Kronecker
product has size $m_Bm_Cm_D \times n_Bn_Cn_D$, and an arbitrary
element of $B \otimes C \otimes D$ can be denoted  $(B \otimes C
\otimes D)_{abc,def}=B_{ad}C_{be}D_{cf}$, where the index written as
$abc$ denotes the number $c+bm_D+am_Cm_D$ and the index $def$ is the
number $f+en_D+dn_Cn_D$; we suppose that the indices start form
zero: $a=0,\ldots,m_B-1$, etc. If $B,C,D$ happen to be square
matrices of size $n$, this notation coincides with the usual
notation in base $n$ of an index running from $0$ to $n^3-1$. This
notation of the Kronecker product is very close to the tensor
product used in algebra and differential geometry. The only
difference is that $B \otimes C \otimes D$, viewed as a tensor
product, is considered as a $6$-dimensional array with $a, b, c, d,
e, f$ as separate indices, instead of a matrix (i.e., a
$2$-dimensional array). All this immediately extends to more than
three matrices.

\subsection{Block Kronecker product}

Let us now consider the following variant of Kronecker product, that
we call \emph{block Kronecker product}. Consider for instance two
matrices $B$ (of size $n^3 \times n^3$) and $C$ (of size $n^2 \times
n^2$). The block Kronecker product of $B$ and $C$ is defined as
follows: its element of index $abcde,ghijk$ is the element
$B_{cde,ghi}C_{ab,jk}$ (notice the shift of the first indices by two
places). We will denote this matrix by $B \odot C$. This definition
applies to any two square matrices whose dimensions are powers of
$n$.  In general, we can write  $(B \odot C)_{p,q}=(B \otimes
C)_{\sigma^t(p),q}$, where $\sigma$ operates a cyclic permutation by
one place to the left on the digits of $p$ in base $n$, and $C$ is
of size $n^t$.

The matrix $M$ defined by Equation (\ref{eq:M}) can be expressed as
$M=(I \otimes \cdots \otimes I) \odot A$ (where the $n \times n$
identity matrix $I$ is repeated $k-1$ times).  If we write the index
of $M$ in base $n$, then $M_{i_{1} \ldots i_k,j_{1} \ldots
j_k}=I_{i_{2},j_{1}}I_{i_{3},j_{2}}\ldots I_{i_{k-1},j_{k}}
A_{i_{1},j_{0}}$.

%%%%An index $i$ on $M$, running from $0$ to $n^l-1$, can be expressed
%%%%in base $n$ as a sequence of $l$ digits, that we denote $i_{l-1}
%%%%\ldots i_0$. Then $M_{i_{l-1} \ldots i_0,j_{l-1} \ldots
%%%%j_0}=I_{i_{l-2},j_{l-1}}I_{i_{l-3},j_{l-2}}\ldots I_{i_0,j_{1}}
%%%%A_{i_{l-1},j_{0}}$.

This form is particularly useful to compute the behavior of $M$ from
the properties of the block Kronecker product, which we now explore.
As a first property, we can easily see that
\begin{equation}\label{eqntranspose}
(B \odot C)^T=C^T \odot B^T.
\end{equation}
We can also prove the following lemma.
\begin{lemma}
For any matrices $A,B,C,D,E,F$ for which all the products below are
meaningful, we have
\begin{equation}\label{lemm11}
  ((A \otimes B) \odot C)( (D \otimes E) \odot F)= BD \odot (CE
\otimes AF).
\end{equation}
\end{lemma}

\begin{proof}
We write, using Einstein's convention (indices repeated twice in an expression
are implicitly
summed over), 
\begin{align*}
&[((A \otimes B) \odot C)( (D \otimes E) \odot F)]_{u,w}=\\
&=((A \otimes
B) \odot C)_{u,v}( (D \otimes E) \odot F)_{v,w}\\
&=A_{u_2,v_1} B_{u_3,v_2} C_{u_1,v_3} D_{v_2,w_1} E_{v_3,w_2}
F_{v_1,w_3},
\end{align*}
where $u$, $v$, and $w$, interpreted as sequences of digits in base
$n$, have been partitioned into $u_1u_2u_3$,  $v_1v_2v_3$, and
$w_1w_2w_3$ in an appropriate way. This is possible if $B$ and $D$
have same size, as well as $C$ and $E$, and $A$ and $F$. Then the
expression above can be regrouped as
\begin{align*}
&(BD)_{u_3,w_1} (CE)_{u_1,w_2} (AF)_{u_2,w_3}=\\
&\qquad\qquad\qquad =(BD \odot (CE \otimes AF))_{u,w},
\end{align*}
which ends the proof.
\end{proof}
In particular, if $B=D=1$ we have
\begin{equation}\label{lemm12}
 (A \odot C )(E \odot F)=(CE
\otimes AF).
\end{equation}
If we choose  $C=E=1$ instead, we have
\begin{equation}\label{lemm13}
 (A \otimes B)(D \odot F)= BD \odot AF.
\end{equation}

The following proposition provides an interesting characterization
of the powers of any order of the matrix $M$.
\begin{proposition} \label{prop1}
For $A$ a square matrix, $M$ defined by Equation (\ref{eq:M}), and
 any integers $r\geq 0$ and
$0\leq s <k$,
$$M^{rk+s}=\underbrace{(A^r \otimes \cdots \otimes A^r)}_{k-s} \odot \underbrace{(A^{r+1}
\otimes
\cdots \otimes A^{r+1})}_s,$$ where the exponents in the right-hand
side sum to $rk+s$.
\end{proposition}

\begin{proof}
We prove the claim by induction on $rk+s$. It is true by definition
for $rk+s=1$. The induction step is easily proved by applying
Equation (\ref{lemm11}). Indeed, $[(A^r \otimes (A^r \otimes \cdots
\otimes A^r)) \odot (A^{r+1} \otimes \cdots \otimes A^{r+1})][((I
\otimes \cdots \otimes I) \otimes (I \otimes \cdots \otimes I))
\odot A )]$ can be written as $(A^r \otimes \cdots \otimes A^r)
\odot (A^{r+1}  \otimes \cdots \otimes A^{r+1} \otimes (A^r A))$.
The argument is correct  also for limit cases $s=0$ and $s=k-1$.
\end{proof}
In particular we have the following.
\begin{corollary} \label{coro}
For $A$ a square matrix and $M$ defined by Equation (\ref{eq:M}),
$$M^k=A \otimes \cdots \otimes A.$$
Moreover, if $A$ satisfies (A),(B),(C) the essential spectral radius of
$M$ is the $k$th root of the essential spectral radius of $A$.
\end{corollary}

\begin{proof}
The first part is a particular case of Proposition \ref{prop1}. From
the properties of Kronecker product, we know the spectrum of $M^k$
is composed of all possible products of $k$ eigenvalues of $A$.
Hence the largest eigenvalue in absolute value, different from $1$,
of the matrix $M^k$ results to be $1^{k-1}\lambda$, where $\lambda$
denotes the largest eigenvalue in absolute value, different from
$1$, of the matrix $A$.
\end{proof}

 This also proves also that conditions (B) and
(C) are verified for $M$ when they are for $A$. If we take
\begin{equation}\label{eqnAdeadbeat} A=1/n\1 \1^T,
\end{equation} of size $n$, then $M^k$ is the matrix $1/n^k \1 \1^T$ of size
$n^k$ with all identical elements. Thus we have a strategy
converging exactly in $k$ steps. We comment further on this example
in the next section. Another property of $M$ that will prove useful
is stated in the next proposition.
\begin{proposition}  \label{prop2}
For $A$ a square matrix,  $M$ defined by Equation (\ref{eq:M}), and
any integers $r\geq 0$ and $0\leq s <k$,
\begin{align*}
&{M^T}^{rk+s} M^{rk+s}=\underbrace{{A^T}^{r} A^r
\otimes \cdots \otimes {A^T}^{r}A^r}_{k-s} \otimes \\
&\qquad\qquad\qquad\qquad\,\,\,\,\, \otimes
\underbrace{{A^T}^{r+1}A^{r+1} \otimes \cdots \otimes
{A^T}^{r+1}A^{r+1}}_s,
\end{align*}
where the sums of exponents is $rk+s$.
\end{proposition}
\begin{proof}
From Proposition \ref{prop1}, we know that $M^{rk+s}=(A^r \otimes
\cdots \otimes
 A^r) \odot (A^{r+1} \otimes \cdots \otimes A^{r+1})$. Hence, by Equation
 (\ref{eqntranspose}), ${M^T}^{rk+s}= ({A^T}^{r+1}
  \otimes \cdots \otimes {A^T}^{r+1}) \odot (A^r \otimes \cdots \otimes A^r)$.
These two expressions are multiplied using Equation (\ref{lemm12}).
\end{proof}
Now we would like to compute $\Tr {M^T}^{t}M^{t+1}$. This will be
useful later when we will evaluate the performance of the block
Kronecker strategy. We first need the following lemma.
\begin{lemma} \label{lemmtrace}
Let $B_0, B_1, \ldots, B_{k-1}$ be $k$ square matrices of same
dimensions. If $l \leq k$ is relatively prime  to $k$, then
\begin{align*}\label{eqnlemmtrace}
&\Tr \, (B_0 \otimes B_1 \otimes \cdots \otimes B_{l-1})
\odot (B_l \otimes \cdots \otimes B_{k-1})=\\
&\qquad\qquad\qquad\Tr \, B_0 B_l B_{2l} B_{3l} \cdots B_{(k-1)l},
\end{align*}
where the indices are understood modulo $k$.
\end{lemma}
\begin{proof}
If we use Einstein's convention (repeated indices are summed over),
we can write
\begin{eqnarray*}
 & & \Tr (B_0 \otimes B_1 \otimes \cdots \otimes B_{l-1}) \odot
(B_l \otimes \ldots \otimes B_{k-1})\\
   &=& [(B_0 \otimes B_1 \otimes \cdots \otimes B_{l-1})
  \odot (B_l \otimes \cdots \otimes B_{k-1})]_{p,p}\\
  &=&  (B_0)_{p_{k-l},p_0} (B_1)_{p_{k-l+1},p_1} \cdots
  (B_{l-1})_{p_{k-1},p_{l-1}} \\
  & & \quad \quad  (B_l)_{p_0,p_l} \cdots (B_k)_{p_{k-l-1},k}\\
  &=& (B_0)_{p_{k-l},p_0} (B_l)_{p_0,p_l} (B_{2l})_{p_l,p_{2l}} \\
  & & \quad \quad (B_{3l})_{p_{2l},p_{l}} \cdots (B_{(k-1)l})_{p_{(k-2)l},p_{(k-1)l}}\\
  &=& \Tr B_0 B_l B_{2l} \ldots B_{(k-1)l},
  \end{eqnarray*}
 where $p=p_0 p_1 \ldots p_{k-1}$.
\end{proof}
\begin{proposition} \label{prop3}
For $A$ and $M$ as defined above, if $A$ is normal (i.e.,
$A^TA=AA^T$) then
$$
\Tr{M^T}^{t} M^{t+1}=\Tr {A^T}^tA^{t+1}.
$$
\end{proposition}
\begin{proof}
We know that ${M^T}^tM^t={A^T}^r A^r \otimes \ldots \otimes
{A^T}^{r+1} A^{r+1}$, if $t=rk+s$ for some $0 \leq s < k$. Hence
$${M^T}^t M^{t+1}=({A^T}^r A^r \otimes \cdots \otimes {A^T}^{r+1}
A^{r+1})((I \otimes \cdots  \otimes I) \odot A),$$ which by Equation
(\ref{lemm13}) is equal to $({A^T}^rA^r \otimes \cdots \otimes
{A^T}^{r+1}A^{r+1})   \odot  {A^T}^r A^{r+1}$. By Lemma
\ref{lemmtrace}, this matrix has the same trace as ${A^T}^rA^r
\ldots {A^T}^{r+1}A^{r+1} {A^T}^r A^{r+1}$. As $A^T$ and $A$
commute, this is also the trace of ${A^T}^{t}A^{t+1}$.
\end{proof}

\subsection{De Bruijn's graph}

The communication graph of $M$ is (a subgraph of) a de Bruijn graph,
which has $n^k$ vertices and arcs from any $i$ to $ni,ni+1, ni+2,
\ldots$ and $ni+k-1$ (all modulo $n^k$). In particular, if $A$ is
given by Equation (\ref{eqnAdeadbeat}), then $M$ is the adjacency
matrix of a de Bruijn graph, normalized so as for every row to sum
to one. This graph was introduced by de Bruijn \cite{deBruijn46} in
1946 and has been considered for efficient distribution of
information in different context such as in parallel computing
\cite{Samathan89} and peer-to-peer networks \cite{FraigniaudG06}.
This paper can be seen as an extension of this idea to consensus
problems.

\subsection{Design decentralisation}
The process itself of convergence to consensus is decentralised,
in the sense that every agent acts on its own.  However the communication
strategy (who talks to whom?) must be designed once for all beforehand. This
can be done in centralised way, where a new external agent dispatch to every
other agent their own strategy.
This can also be done in a decentralised way, where every agent is attributed
a number $i$ between $0$ and $N-1$ and then finds the agents of number $\nu
i$, $\nu i +1$,\ldots, $\nu i +\nu-1$. Achieving this in the most  effective
way is a problem of its own, and is not treated in this paper.

\section{The quickest possible strategy}
%JCD: NEW SUBSECTION

We have seen that starting from $A$ with all identical entries, we
get arbitrarily large matrices $M$ converging in finite time $k$. If
we write $N=n^k$ the dimension of $M$, this convergence time is
$\log N/\log n=\log N/\log \nu$, where $\nu$ is the maximal
in-degree of the graph of communication for $M$. We can see that no
strategy, whether linear or not, whether time-invariant or not, can
converge more rapidly. Indeed, to reach the average of the initial
conditions, every agent must have information about all other
agents, but it can only know $\nu$ other positions in one step of
time, $\nu^2$ in two steps of time, etc. Hence the propagation of
information needs around $\log N / \log \nu$ steps to connect all
agents. This reasoning is made rigorous in the following
proposition.
\begin{proposition}
Let $M\,\in\,\R^{N \times N}$ such that $M \geq 0$. Let $\nu$ be
defined as above. Then $M^k>0$ implies $\nu^k\geq N.$
\end{proposition}

\begin{proof}
The fact $M^k>0$ implies that for any pair of nodes $(i,j)$ there
exists in the graph $\mathcal{G}_M$ a path connecting $i$ to $j$ of
length $k$. Hence there are at least $N^2$ paths of length $k$. Let
now $P_i$ denote the number of paths having length $i$. The previous
consideration implies that $P_k \geq N^2$. On the other hand it is
easy to see that $P_1\leq \nu N$ and in general that $P_i \leq \nu^i
N$ from which we get that $P_k\leq \nu^k N$. Hence $\nu^k N \geq
N^2$ from which it results that $\nu^k \geq N$.
\end{proof}

The above proposition considers only the time-invariant case. An
identical result can be found for the time-varying case, showing
that there is no difference, in terms of speed of convergence toward
the meeting point, between the time-invariant and the time-varying
cases. This can be seen an \emph{a posteriori} justification of our
interest in the class of the time-invariant strategies.

A linear time-invariant strategy converges in finite time iff its
essential spectral radius  is $0$.  For a strategy converging in
infinite time, the essential spectral radius is a good measure of
the convergence to the average of the initial conditions, as already
mentioned.

\subsection{Comparison between block Kronecker strategy and Cayley strategy}

In this subsection we propose a comparison of the block Kronecker
strategy with another strategy whose underlying communication graph
has limited max in-degree and exhibits strong symmetries: the Cayley
strategy.\\
First we recall the concept of Cayley graph defined on Abelian
groups~\cite{Babai79-1,Alon94-1}. Let $G$ be any finite Abelian
group (internal operation will always be denoted $+$) of order
$|G|=N$, and let $S$ be a subset of $G$ containing zero. The Cayley
graph ${\cal G}(G,S)$ is the directed graph with vertex set $G$ and
arc set
$$
         \mathcal{E}=\left\{(g,h) : h-g \in S
         \right\}.
$$
Notice that a Cayley graph is always in-regular, namely the
in-degree of each vertex is equal to $|S|$. Notice also that strong
connectivity can be checked algebraically. Indeed, it can be seen
that a Cayley graph ${\cal G}(G,S)$ is strongly connected if and
only if the set $S$ generates the group $G$, which means that any
element in $G$ can be expressed as a finite sum of (not necessarily
distinct) elements in $S$. If $S$ is such that $-S=S$ we say that
$S$ is inverse-closed. In this case the graph obtained is
undirected.

A notion of Cayley structure can also be introduced for matrices.
Let $G$ be any finite Abelian group of order $|G|=N$. A matrix
$P\in\R^{G\times G}$ is said to be a Cayley matrix over the group
$G$ if
$$P_{i,j}=P_{i+h,j+h}\qquad\forall\ i,j,h\in G\,.$$
It is clear that for a Cayley matrix $P$ there exists a $\pi:G\to
\R$ such that $P_{i,j}=\pi(i-j)$. The function $\pi$ is called the
generator of the Cayley matrix $P$. Notice how, if $P$ is a Cayley
matrix generated by $\pi$, then ${\cal G}_P$ is a Cayley graph with
$S=\{h\in G:\pi(h)\not=0\}$. Moreover, it is easy to see that for
any Cayley matrix $P$ we have that $P\1=\1$ if and only if $\1^T
P=\1^T$. This implies that a Cayley stochastic matrix is
automatically doubly stochastic. In this case the function $\pi$
associated with the matrix $P$ is a probability distribution on the
group $G$. Among the multiple possible choices of the probability
distribution $\pi$, one is particularly simple, namely
$\pi(g)=1/|S|$ for every $g\in S$.
\begin{example}
\label{ex:weighted_cayley_graph} Let us consider the group $\Z_N$ of
integers modulo $N$ and the Cayley graph ${\cal G}(\Z_N,S)$ where
$S=\{-1,0,1\}$. Notice that in this case $S$ is inverse-closed.
Consider the uniform probability distribution
$$\pi(0)=\pi(1)=\pi(-1)=1/3$$
The corresponding Cayley stochastic matrix is given by
\begin{equation}\label{P:example1}
        P = \begin{bmatrix}
                 1/3  & 1/3 & 0 & 0 & \cdots  & 0 &1/3\\
                 1/3 & 1/3 & 1/3 & 0 & \cdots  & 0 & 0\\
                 0 & 1/3 & 1/3 & 1/3 &  \cdots  & 0 & 0\\
                 \vdots & \vdots & \vdots & \vdots & \cdots & \vdots&
                 \vdots\\
                 1/3& 0 & 0 & 0 & \cdots  & 1/3 & 1/3
            \end{bmatrix}\,.
\end{equation}
Notice that in this case we have two symmetries. The first is that
the graph is undirected and the second that the graph is circulant.
These symmetries can be seen in the structure of the transition
matrix $P$ that, indeed, turns out to be both symmetric and
circulant \cite{Davis79}.
\end{example}

\begin{example}
Let us now consider the group $\Z_N\times \Z_N$ and the Cayley graph
${\cal G}(\Z_N\times \Z_N,S)$ where $S=\{(1,0);(0,0);(0,1)\}$. Again
consider the uniform probability distribution
$$\pi((0,0))=\pi((1,0))=\pi((0,1))=1/3$$
The corresponding Cayley stochastic matrix is given by the following
block circulant matrix belonging to $\R^{N^2 \times N^2}$
\begin{equation}\label{P:example2}
        P = \begin{bmatrix}
                 P_1  & P_2 & 0 & 0 & \cdots & 0 & 0 &0\\
                 0 & P_1 & P_2 & 0 & \cdots & 0 & 0 & 0\\
                 0 & 0 & P_1 & P_2 &  \cdots & 0 & 0 & 0\\
                 \vdots & \vdots & \vdots & \vdots & \cdots & \vdots& \vdots&
                 \vdots\\
                 P_2& 0 & 0 & 0 & \cdots & 0 & 0 & P_1
            \end{bmatrix}
\end{equation}
where $P_1,\,P_2\,\in\R^{N\times N}$ are such that
\begin{equation}\label{P:example2}
        P_1 = \begin{bmatrix}
                 1/3  & 1/3 & 0 & \cdots  & 0 &0\\
                 0 & 1/3 & 1/3 & \cdots  & 0 & 0\\
                 \vdots & \vdots &\vdots & \cdots & \vdots&
                 \vdots\\
                 1/3& 0 & 0& \cdots & 0 & 1/3
            \end{bmatrix},\,
P_2=\frac{1}{3}I.
\end{equation}
This example can be generalized to the more general case of the
discrete $d$-dimensional tori $\Z_N^d$, extensively studied in the
literature regarding the peer-to-peer networks \cite{Ratnasamy},
\cite{Stoica}.
\end{example}

Now we recall an interesting result regarding the essential spectral
radius of the Cayley stochastic matrices. Assume that
$P\,\in\,\R^{N\times N}$ is a Cayley stochastic matrix generated by
a suitable $\pi$ and assume that $|S|=\nu$, where $S$ is as
previously defined. Moreover assume that $0\,\in\,S$. Notice that
this last fact implies that $P_{ii}>0,\,\forall\,i:\,1\leq i \leq
N.$ Then it follows that $\rho \geq 1-C/N^{2/(\nu-1)}$, where $C>0$
is a constant independent of $S$ and $N$ the number of agents. This
result was proved in \cite{Zampieri05}.

On the other side, the block Kronecker strategy constructed from any
matrix $A$ has essential spectral radius $|\lambda|^{1/k}$, where
$|\lambda|$ is the essential spectral radius of $A$, as stated in
Corollary \ref{coro}.

If $0<|\lambda|<1$, then $|\lambda|^{1/k}$ behaves like $1-\mu/k$
for large $k$ and some $\mu$. Recall that $k$ is $\log N/\log n$.
Hence this
is better than abelian Cayley strategies.    %%%%%JC: TO KEEP? if $\nu >2$.

In conclusion, block Kronecker strategies have a better essential
spectral radius, hence a quicker convergence speed, than Cayley
strategies. For the particular choice given by Equation
(\ref{eqnAdeadbeat}), we converge in finite time, and this time is
the smallest possible over all linear strategies with the same
constrained degree.

\subsection{Simulation result}

\begin{figure}[!ht]
\centering
\includegraphics[width=0.45\hsize]{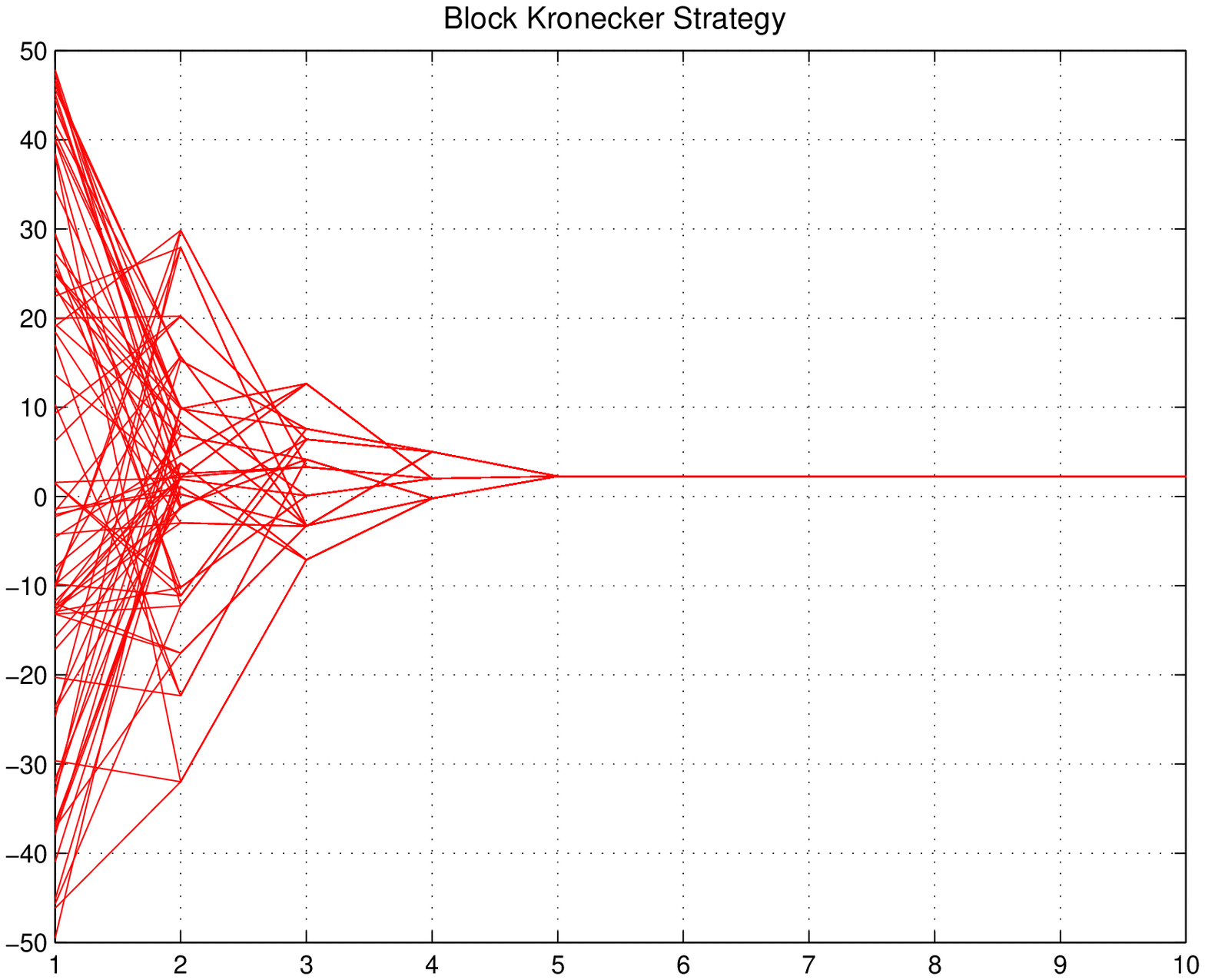}
\includegraphics[width=0.45\hsize]{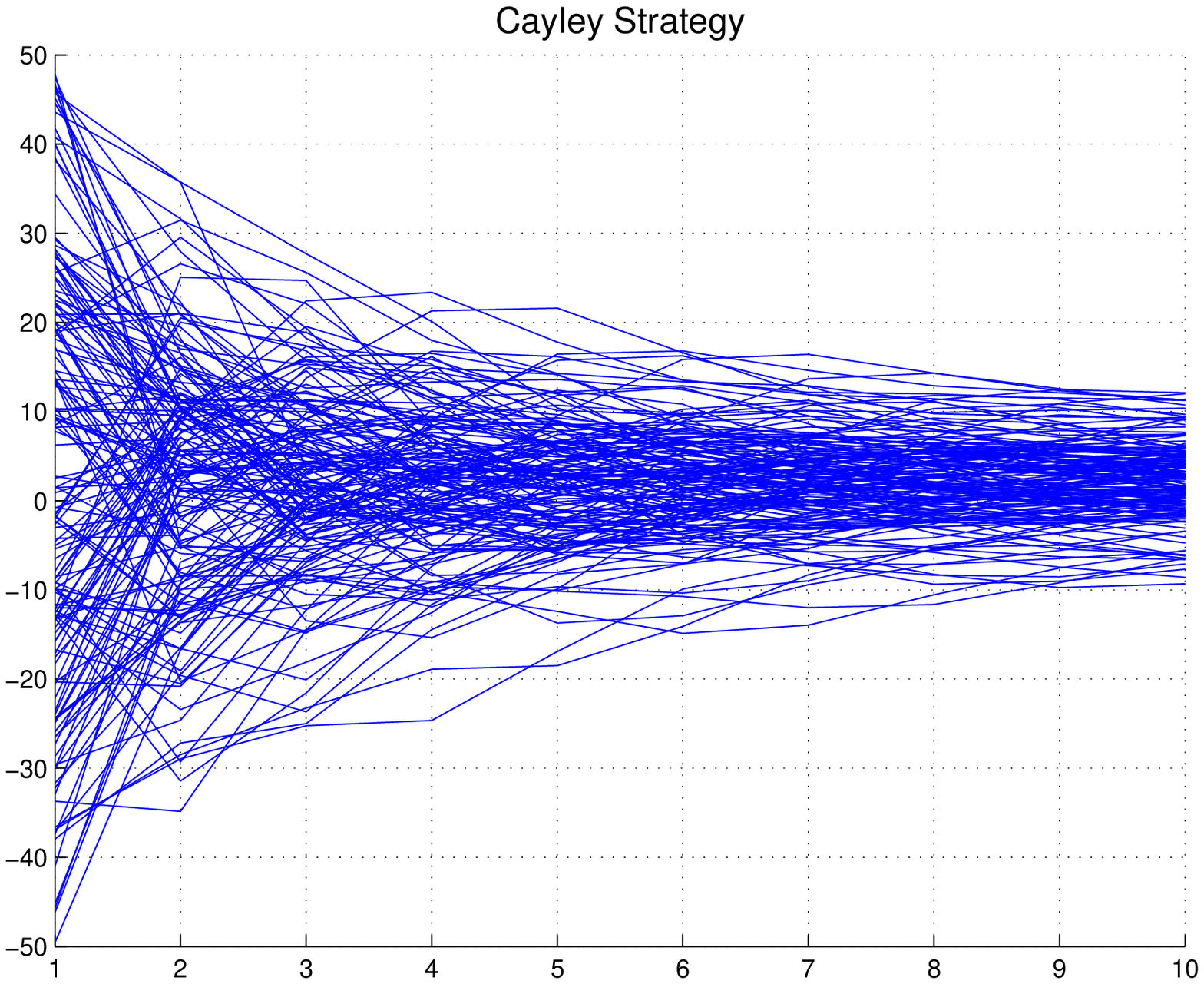}
    \caption{The block Kronecker strategy (left) converges in finite time,
    while the Cayley strategy (right) has a relatively slow convergence}\label{fig:Kron}
\end{figure}

As an illustration, we present a simulative comparison between the
Cayley strategy and the block Kronecker strategy. The network
considered consists of $N=81$ agents. The matrix $P$ for the Cayley
strategy is the matrix (\ref{P:example1}), whereas the matrix $M$
for the block Kronecker strategy is built starting from
$$
A=\left[
\begin{array}{rcl}
1/3 & 1/3 & 1/3 \\
1/3 & 1/3 & 1/3 \\
1/3 & 1/3 & 1/3
\end{array}
\right]
$$
with $n=3$ and $k=4$. The initial conditions has been chosen
randomly inside the interval $[-50,50]$. In both cases the in-degree
is $3$. Notice that, as depicted in Figure \ref{fig:Kron}, 
%and Figure \ref{fig:Cay} %JCD: useless because figures merged
the block Kronecker strategy reaches the average of
the initial conditions in a finite number of steps whereas, the
Cayley strategy, after the same numbers of steps, is still far from
converging toward the meeting point.

\section{LQR cost}

In this section we want to evaluate the performance of the block
Kronecker strategy according to the quadratic cost $J=J_1 + \gamma
J_2$, where $J_1= \mathbb{E}\sum_{t \geq 0} (x(t)-x(\infty))^T
(x(t)-x(\infty))$ accounts for the quickness of convergence,
$J_2=\mathbb{E} \sum_{t \geq 0} (x(t+1)-x(t))^T(x(t+1)-x(t))$ limits
the norm of the updates, and $\gamma $ weights the respective
importance of those two factors. Precisely we evaluate $J$ for any
block Kronecker strategy derived from a normal matrix $A$. Remember
that the initial state $x(0)$ is supposed to be characterized by a
identity covariance matrix. We start with a lemma which provides an
upper and a lower bound for $J_1$.
\begin{lemma} \label{lemm:J1}
If $A$ is a normal $n \times n$ matrix satisfying conditions
(A),(B),(C), and $\rho$ is the  essential spectral radius of $A$,
then the $J_1$ cost of the corresponding block Kronecker strategy of
size $n^k$ satisfies:
$$
J_L\leq J \leq J_U,
$$
where $J_L=N \frac{1- (\Tr (A^TA)/n)^k}{1- \Tr (A^TA)/n}-k$ and
$J_U= J_L+\frac{k}{1-\rho^2}(\Tr A^TA-1).$
\end{lemma}
\begin{proof}
Classical arguments lead to write: \begin{eqnarray*}
J_1 &=& \mathbb{E}\sum_{t \geq 0} (x(t)-x(\infty))^T (x(t)-x(\infty)) \\
    &=& \sum_{t \geq 0} \mathbb{E} \,\,\, (x(t)-x(\infty))^T (x(t)-x(\infty))\\
    &=& \sum_{t \geq 0} \mathbb{E} \,\,\, \Tr (x(t)-x(\infty))^T (x(t)-x(\infty))\\
    &=&  \sum_{t \geq 0}  \mathbb{E} \,\,\, \Tr (x(t)-x(\infty)) (x(t)-x(\infty))^T\\
    &=&  \sum_{t \geq 0}  \Tr (M^t-E) \mathbb{E} (x(0) x(0)^T) (M^t-E)^T\\
    &=&  \sum_{t \geq 0} \Tr (M^t-E) (M^t-E)^T\\
\end{eqnarray*}
with $E=1/n^k \1 \1^T$.
%with $E=1/n^k [1 \quad 1 \ldots 1] [1 \quad 1 \ldots 1]^T$.

Now, $\Tr (M^t-E) (M^t-E)^T=\Tr M^t {M^t}^T -\Tr
 E=\Tr M^t {M^t}^T -1$. When $M$ is derived by block Kronecker product
from a normal matrix $A$, this is equal to $(\Tr (A^TA)^r)^{k-s}(\Tr
(A^TA)^{r+1})^{s}$ if $t=rk+s$, according to Proposition
\ref{prop2}.

We get a lower bound on $J_1$ by summing only the first $k$
terms:

\begin{eqnarray}
 J_1 &=&   \sum_{r \geq 0} \sum_{s=0}^{k-1}  ((\Tr (A^TA)^r)^{k-s} (\Tr (A^TA)^{r+1})^{s}-1)
 \nonumber\\
 &\geq&   \sum_{s=0}^{k-1} (\Tr (A^TA)^0)^{k-s}  (\Tr (A^TA)^1)^{s}-k  \label{eq:lowboundJ1}\\
 &=& \sum_{s=0}^{k-1} n^{k-s}  (\Tr A^TA)^{s}-k \nonumber,
% %
%J_1 &=& \sum_{r \geq 0} \sum_{s=0}^{k-1} ((1+\sum_i \lambda_i^t)^r  (1+\sum_i
%%\lambda_i^{t+1})^{t+1})^{k-r}-1) \\
%&\geq&   \sum_{r=0}^{k} (1+\sum_i \lambda_i^0)^r  (1+\sum_i \lambda_i^{1})^{k-r}-(k+1)
\end{eqnarray}
The last summation is a geometric series that can be evaluated,
leading to the bound
$$
J_1 \geq N \frac{1-(\Tr A^TA/n)^k}{1-\Tr A^TA/n} -k
$$
This proves the left inequality in the claim.

For the right inequality, we find an upper bound on the terms
neglected in the lower bound (\ref{eq:lowboundJ1}). As normal
matrices can be diagonalized by a unitary transformation, the
eigenvalues of $A^TA$, which we denote $1,\lambda_1, \lambda_2,
\ldots, \lambda_{n-1}$, are precisely the squared module of the
eigenvalues of $A$. In particular, $\rho^2=\lambda_1$, and the trace
of $(A^TA)^t$ is  $1+\sum \lambda_i^t$.

The terms neglected in the lower bound (\ref{eq:lowboundJ1}) are

$$\sum_{r \geq 1} \sum_{s=0}^{k-1}((1+\sum_i \lambda_i^r)^{k-s}  (1+\sum_i
\lambda_i^{r+1})^{s}.$$

For every $r$, we bound every of the $k$ terms by the highest (for
which $s=0$). Hence the neglected terms are bounded from above by:
\begin{eqnarray*}
\sum_{r \geq 1} ((1+\sum_i \lambda_i^r)^k -1)&=& k \sum_{r \geq 1}
P(\lambda_1^r,\ldots,\lambda_{n-1}^r),
\end{eqnarray*}
where $P$ is a polynomial in the variables $\lambda_1, \ldots,
,\lambda_{n-1}$ with no independent term: all monomials have degree
at least one. Now we can sum all corresponding monomials for
$r=1,2,\ldots$: this is a geometric series of progression at most
$\lambda_1$. Hence $\sum_{r \geq 1}
P(\lambda_1^r,\ldots,\lambda_k^r)$ is at most $\frac{1}{1-\lambda_1}
P(\lambda_1,\ldots,\lambda_k)=\frac{1}{1-\lambda_1} (\Tr A^TA - 1)$.

Hence $J_1$ differs from our lower bound by at most
$k\frac{1}{1-\lambda_1}(\Tr A^TA - 1)$.\end{proof}

Thus $J_1=N \frac{1- (\Tr (A^TA)/n)^k}{1- \Tr (A^TA)/n} +
\mathcal{O}(\log N)$. Now we estimate $J_2$.

\begin{lemma}
Under the assumptions of Lemma \ref{lemm:J1}, if $\rho_i$ denote the
eigenvalues of $A$ different from one,
$$2 J_1 -N - \sum_i \frac{1}{1- |\rho_i|^2} \leq J_2 \leq 2 J_1 - N.$$
\end{lemma}
\begin{proof}
First we notice, adapting the first steps of the proof of the
preceding proposition, that $J_2= \sum \Tr(M-I)^T(M^T)^tM^t(M-I)$,
with $I$ the identity. This involves terms of the form
$(M^T)^{t+1}M^t$. More precisely,

 \begin{align*}
J_2&=\sum_{t \geq 0} \Tr({M^T}^{t+1}M^{t+1}-{M^T}^{t+1}M^t-\\
&\qquad\qquad-{M^T}^tM^{t+1}+{M^T}^tM^t)\\
&=2 \sum_{t \geq 0} (\Tr {M^T}^tM^t -1) - N -\\
  &\qquad\qquad -2 \sum_{t \geq 0}
  (\Tr{M^T}^tM^{t+1} -1)
\end{align*}

The first term  of the last member  is precisely $2 J_1$, the last
term is, thanks to Proposition \ref{prop3}, $2 \sum_{t \geq 0}
  (\Tr{A^T}^tA^{t+1} -1)$.

From Cauchy-Schwartz inequality applied to Frobenius norm, $\Tr
{A^T}^tA^{t+1} \leq \sqrt{\Tr {A^T}^{t+1}A^{t+1}  \Tr {A^T}^{t}A^{t}
} \leq \Tr {A^T}^{t}A^{t}$.

Hence $\sum_{t \geq 0} (\Tr {A^T}^tA^{t+1} -1) \leq \sum_{t \geq 0}
\sum_i \lambda_i^t = \sum_i \frac{1}{1- \lambda_i}$, where, as argued
in the proof of Lemma \ref{lemm:J1}, $\lambda_i=|\rho_i|^2$.
%
%The rightmost member does not comprise the terms of the form
%$(M^T)^{t+1}M^t$, while the the leftmost does. The estimation is
%done by a Cauchy-Schwartz inequality of the Frobenius norm of
%$(A^T)^{t} A^{t+1}$. To be detailed.
\end{proof}

Hence, $$J =  N \Big((1+2\gamma)\frac{1- (\Tr (A^TA)/n)^k}{1- \Tr
(A^TA)/n} -\gamma+\mathcal{O}(\log N/N)\Big).$$ Since the trace of
$A^TA$ is the sum of squares of elements of $A$,  we see that the
coefficient of $N$ (neglecting the $\mathcal{O}(\log N/N))$ term) is
optimized by the matrix $A=1/n \1\1^T$, whatever the value of
$\gamma$ is. In this case, the lower bound obtained on $J_1$ is
exact, since only $k$ terms are non-zero. The optimal cost is then
$$
J=N\Big((1+2\gamma)\frac{1-1/N}{1-1/n}-\gamma +\mathcal{O}(\log
N/N)\Big),
$$
with $\nu=n$.

Hence there is here no trade-off between $J_1$ and $J_2$ among the
family of block Kronecker strategies, in contrast with the general LQR
theory.

Note that the optimal control strategy for unconstrained degree
(every agent knows every position) is easily solved by a scalar
algebraic Riccati equation, leading to the optimal cost $J=N
(1+\sqrt{1+4\gamma })/2$. If $\gamma$ is small and $n$ is large,
then the optimal finite-time block Kronecker approaches the
unbounded degree optimal solution with a cost approximately equal to
$(1+\gamma)N$.

\section{Conclusions}
We have introduced a family of strategies for a consensus problem,
whose graph of communication is de Bruijn's graph. We have shown
that they can converge in finite logarithmic time, which is optimal.
We have evaluated the LQR cost of these strategies, proving their
quasi-optimality if the cost of update is small and the degree of
the graph not too low.

This work can be extended in several directions, including:

\begin{itemize}

\item designing strategies valid for any $N$, not only exact powers of
$n$;

\item tackling the continuous-time case, where no deadbeat strategy can exist;

\item estimating the LQR cost for Cayley strategies;

\item finding strategies that minimize the LQR cost for any cost $\gamma$
of the update;

\end{itemize}

\section{Acknowledgements}
This work was partly developed during the stay of one of the authors
(J.-C. D.) at Department of Information Engineering of Universit\`a
di Padova. Stimulating discussions with J. Hendrickx are gratefully
acknowledged.

\bibliographystyle{abbrv}

%\bibliography{consensus1}
%

\end{document}